\documentclass[3p,times]{elsarticle}

\usepackage{ecrc}

\volume{00}

\firstpage{1}

\journalname{Physics Procedia}

\runauth{}

\jid{procs}

\jnltitlelogo{Physics Procedia}

\CopyrightLine{2014}{Published by Elsevier Ltd.}

\usepackage{amssymb}
\usepackage{graphicx}
\usepackage[figuresright]{rotating}

\begin{document}

\begin{frontmatter}



\dochead{}
%% e.g. \dochead{17th International Conference on Dynamical Processes in Excited States of Solids}

\title{Geo-neutrinos and Earth Models}


\author[label1]{S.T. Dye}
\author[label2]{Y. Huang}
\author[label2]{V. Lekic}
\author[label2]{W.F. McDonough}
\author[label2]{O. \v Sr\'amek}

\address[label1]{Hawaii Pacific University, Kaneohe, HI 96744 U.S.A.}
\address[label2]{University of Maryland, College Park, MD 20742 U.S.A.}

\begin{abstract}
We present the current status of geo-neutrino measurements and their implications for radiogenic heating in the mantle. Earth models predict different levels of radiogenic heating and, therefore, different geo-neutrino fluxes from the mantle. Seismic tomography reveals features in the deep mantle possibly correlated with radiogenic heating and causing spatial variations in the mantle geo-neutrino flux at the Earth surface. An ocean-based observatory offers the greatest sensitivity to the mantle flux and potential for resolving Earth models and mantle features. Refinements to estimates of the geo-neutrino flux from continental crust reduce uncertainty in measurements of the mantle flux, especially measurements from land-based observatories. These refinements enable the resolution of Earth models using the combined measurements from multiple continental observatories.
\end{abstract}


\end{frontmatter}

%\linenumbers

\section{Introduction}
Geo-neutrinos stream to space from the decay of radioactive isotopes within the Earth. Their flux at the surface is predicted to vary spatially, primarily due to variations in continental crust - in which radioactive isotopes are concentrated - with secondary effects due to possible compositional heterogeneities in the mantle. Geo-neutrino flux measurements, which account for the predicted variation, estimate radiogenic heating in the mantle. This heating helps power plate tectonics, earthquakes, volcanism, and mantle convection. Estimates of radiogenic heating in the mantle constrain Earth models, providing important information on the origin and thermal history of the planet.
\par
Ongoing measurements of the surface flux of geo-neutrinos, which are initiated by the decays of $^{238}$U and $^{232}$Th, provide an encouraging outlook for resolving mantle heating. At Japan the flux measurement is $(3.4\pm0.8)\times10^6$ cm$^{-2}$ s$^{-1}$ \cite{gando13}, while at Italy the flux measurement is $(4.3\pm1.3)\times10^6$ cm$^{-2}$ s$^{-1}$ \cite{bellini13}. Subtracting the predicted flux from the crust \cite{enomoto, coltorti, huang13} and assuming negligible flux from the core estimates the flux from the mantle. It is customary to express geo-neutrino flux as a rate of recorded interactions in a perfect detector with a given exposure. The usual unit is the terrestrial neutrino unit (TNU) \cite{mantovani}.
\par
\begin{figure}[h!]
\centering
\includegraphics[width=5in]{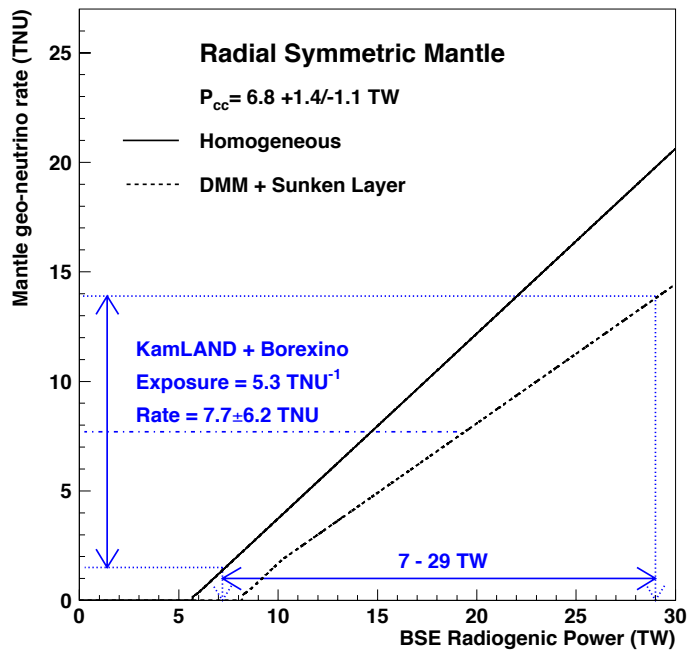}
\caption{\label{BSE}Radiogenic power of bulk silicate Earth (BSE) is constrained between 7 and 29 TW by the combined measurement of geo-neutrinos from the mantle \cite{luza13}. The lower bound on BSE radiogenic heating obtains with a homogeneous mantle and the lower limit of radiogenic heating in continental crust. The upper bound on BSE radiogenic heating obtains with a 150-km thick sunken layer enriched in heat-producing elements at the base of depleted MORB-source mantle and the upper limit of radiogenic heating in continental crust.}
\end{figure}

Combining the surface geo-neutrino flux measurements at Japan and Italy estimates a mantle geo-neutrino detection rate of $7.7 \pm 6.2$ TNU \cite{luza13}.  The main sources of uncertainty in the estimate are the experimental errors in the flux measurements and limited knowledge of the subtracted crust fluxes \cite{enomoto, coltorti, huang13}. Translating the detection rate into radiogenic heating introduces additional uncertainty due to ambiguity in the amounts and distributions of uranium and thorium in the mantle \cite{dye10, dye12}. Standard assumptions for the ratios of thorium to uranium and of potassium to uranium \cite{arevalo09} suggest $2 - 21$ TW of radiogenic heating in the mantle (Figure \ref{BSE}). 

\par
Radiogenic heating in the mantle is an important component of the energy budget of the Earth. The other main components are the flow of heat across the core-mantle boundary \cite{lay08} and the rate at which the mantle sheds primordial heat. Together with the heat generated in the crust, these sum to the surface heat flux of $47 \pm 3$ TW \cite{davies10}. Improving the estimate of mantle radiogenic heating better constrains models of the origin and thermal evolution of the Earth. Although the required precision is currently under investigation, there is clearly much room for improvement in this estimate. Additional exposure to the flux of geo-neutrinos reduces experimental errors and detailed investigations of the crust better defines the subtracted crust fluxes \cite{huang13}. These improve precision but do not eliminate uncertainty in the radiogenic heating. Ambiguity in the amounts and distributions of uranium and thorium in the mantle remain. For example, a perfect measurement of the mantle signal rate at the central value of the estimated $7.7$ TNU suggests $8 - 11$ TW of radiogenic heating in the mantle, assuming radial symmetry. Assigning increased uranium and thorium concentrations to deep mantle seismic structures suggests a similar spread in values and motivates observational strategies \cite{sramek13}.
\par 
We describe various Earth models classified by low, medium, and high levels of radiogenic heating. These models all accommodate the estimated 7 TW of radiogenic heating in the crust \cite{huang13}. Depending on the model the residual excess in the mantle is as low as 3 TW \cite{javoy10} and as much as 18 TW. In addition to minimum levels of mantle radiogenic heating defined by studies of the source of mid-ocean ridge basalt (MORB) \cite{work, salt, arev10}, we consider heterogeneity of radioactive isotopes associated with seismically resolved mantle structures, including large low seismic velocity provinces (LLSVPs) and ultra-low velocity zones (ULVZs). Variation in the values of the radioactive isotope content of MORB-source mantle and potential heterogeneity of radioactive isotopes associated with seismic structures contribute uncertainty to the estimate of mantle radiogenic heating. We then describe observational strategies for reducing these uncertainties and resolving mantle models.
\section{Earth Models}
Models developed to explain the composition of the Earth and the abundances of the heat producing elements (i.e., K, Th and U) therein are developed from concepts that use chondritic meteorites, which are primitive undifferentiated solar system materials assembled during the initial accretion and formation of the star and planets, to build the Earth. These models are required to be consistent with physical and chemical observables of the crust, mantle, core system.  The families of compositional models that describe the Earth span a factor of three in the abundances of the heat producing elements and  differences are due to contrasts in starting assumptions. In addition, some models shape composition by invoking major post-accretionary processes, such as significant mass losses early in the differentiation and evolution of the young planet during the waning stages of accretion. The present-day structure of the Earth is well defined from the one-dimensional seismic profile of the Earth (PREM, Preliminary Reference Earth Model) \cite{dziand81}, which includes, from surface to center, a thin continental ($\sim35$ km) and oceanic ($\sim8$ km) crust surrounding a thick mantle ($\sim2900$ km), both of which are made up of silicates, and a Fe-Ni alloy core ($\sim3500$ km), which also contains one or more significant light elements (Z $< 26$) to compensate for its reduced density.
\par
The interior of the Earth is characterized by a series of seismic discontinuities that identify boundaries between major regions of the Earth and define profiles of density with depth. Given insights from mineral physics on the elastic properties of materials at high pressures and temperatures, these profiles can be used to estimate the major element composition of the planet with depth. The major seismic discontinuities seen in the bulk silicate Earth (BSE, present day crust and mantle) are associated with compositional changes (e.g., the Moho boundary between the crust and the mantle) or are either solely mineral phase changes or both phase and compositional changes as those observed at 410 km and 660 km depth.  The composition of the Upper Mantle (beneath the Moho and above the 410 km seismic boundary) is widely accepted, given it is readily sampled by basalt (derivative material) and peridotite (direct mantle samples brought to the surface by tectonic forces), and this region of the mantle contains olivine (55 \%), two pyroxenes (ortho and clino types) and an aluminous phase (plagioclase, spinel, or garnet depending on the pressure).  There is considerable discussion on the composition of the Transition Zone (between the 410 and 660 km discontinuities) and the Lower Mantle (between the 660 km discontinuity and the core-mantle boundary), with some models predicting a gross compositional boundary at 660 km, with the Lower Mantle having a convection mode separate from that of the Upper Mantle and Transition Zone \citep [e.g.][]{javoy10}.  Alternative models view the seismic discontinuities in the mantle to be strictly phase changes - 410 km marks the iso-chemical change from olivine to wadsleyite and 660 km is where ringwoodite (the higher pressure polymorph of wadsleyite) undergoes an iso-chemical disproportionation to Mg-perovskite and ferropericlase.

\par
\begin{figure}[h!]
\centering
\includegraphics[width=5in]{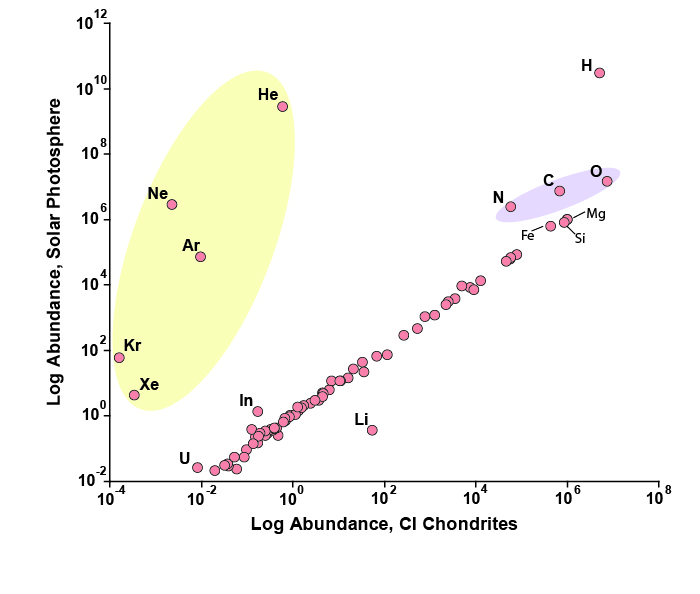}
\caption{\label{Sun_C1}Plot of the elemental abundances in the Sun's surface (photosphere) versus those in CI carbonaceous chondrites. The solar abundances of the highly volatile elements H, C, N, O and the noble gases are distinctly higher than in the chondrite reference, as is the case for rocky materials in the solar system.  The remaining elements follow a one to one ratio, including Fe, Mg, and Si, the abundant rock forming elements.  Solar depletion in lithium is due to its processing the Sun's nuclear burning cycle.  Chondrite depletions in U and In are noted as compared to the solar abundances.}
\end{figure}

There is a non-uniqueness to the relative proportion of Fe, O, Mg and Si in the BSE based on compositional insights from chondritic meteorites, the building blocks of the planets.  On a mass basis, these four elements, however, constitute at least 92 weight \% of the BSE, with the addition of Ca, Al and Ni defining the remaining fraction and describing more than 98 weight \% of its bulk composition \cite{mcdoug14}.  The seismic profile of the Lower Mantle and the known elastic properties of the three minerals that compose the bulk of this region (Mg-perovskite, Ca-perovskite and ferropericlase) also provide non-unique solutions to the relative mode proportions of these phases.  Consequently, some models conclude that the Earth has a C1 carbonaceous chondrite composition (particularly for its Mg/Si value, \citep[e.g.][] {murakami}, while others recognize that the relative and absolute abundances of the four most abundance elements in the Earth vary widely between the varieties of chondrites \citep[e.g.][] {palme14}. The C1 carbonaceous chondrites are volatile rich and have an amazing compositional match (aside from H, C, N, O and noble gases), spanning seven orders of magnitude, with that of the solar photosphere's composition (Figure \ref{Sun_C1}).  Other Earth models match the planet's composition to that of Enstatite chondrites, emphasizing the identical oxygen isotopic compositions of chondrites and the Earth and high iron content \cite{javoy10}.

\par
There is a richness of geophysical, geochemical and meteoritic observables that allow for a range of acceptable parameter space and thus, a non-uniqueness to the potential solutions that describe the composition of the Earth.  This framework of viable models provides a welcomed and much needed environment for the introduction of geo-neutrino constraints on bulk Earth compositional models.  Importantly, if we can describe the absolute concentration of the Th and U content of the BSE, we have defined the absolute amount of refractory elements (elements with condensation temperature greater than 1350 K in a solar nebular environment where the half-mass fraction is incorporated into precipitated phases).  Elements like Ca, Al, Th and U are all refractory elements;  a constraint on one of these elements is a constraint on all, as ratios of these elements are conserved in chondrites.  In contrast to the refractory elements, K is a moderately volatile element and there is no conserved ratio among the chondrites for element pairs like K/U.  Consequently, much effort is invested in understanding variations in elemental ratios like K/U, which for the Earth appears to be relatively constant at $\sim~14,000$ \cite{arevalo09}.  In terms of present-day radiogenic heat production in the Earth there is a continuum of compositional models but overall three families can be identified: the low Q models (10-15 TW of power from K, Th and U), medium Q models (17-22 TW) and high Q models ($>25$ TW), which have been previously referred to as cosmochemical, geochemical, and geodynamical models, respectively \cite{mcd12, sramek13}.
\par
The Low Q models: These models include the Collisional Erosion model \cite{onepal08}, the Enstatite Earth model \cite{javoy10}, and the Non-Chondritic model \cite{camone12}. These models predict that the Earth either accreted with low quantities of Th and U \cite{javoy10} or that in the early stages of accretion the planet lost up to half of its Th and U budget due to accretionary impacts that involved an excess of impact energy and consequently the fiery explosive expulsion of mass from the early Earth, and in this case a differentiated crust enriched in Th and U.
\par
The Medium Q models: These models use samples of the Earth's mantle, with controls from ratios of refractory elements in chondrites, to determine the compositional characteristics of the mantle \citep[e.g.][]{hartz86, mcdsun95, palmeoneill03}. They are very much dependent on the assumption that what you get at the top of the Earth reflects the bulk composition of what is in the mantle.
\par
The High Q models:  These models typically evaluate the Earth's energy from a physics approach with boundary condition given by the surface heat flux and seek solutions to the thermal evolution of the planet by modeling parameterize mantle convection in terms of the force balance \cite{schubert80,  turcotte}. Trade offs between buoyancy and viscosity versus thermal and momentum diffusivities allow for a range of solutions that can include Low to High Q models.

\section{Crust Model Refinements}
The parents of geo-neutrinos, namely U, Th, and K, are all highly incompatible elements (defined as having crystal/melt partition coefficients much less than unity), and thus preferably concentrated in melts relative to residual solid phase during partial melting. The crust and mantle differentiation process leads to the concentrations of these heat producing elements in the continental crust, which itself contributes $\sim7$ TW radiogenic heating. Current KamLAND, Borexino and coming SNO+ geo-neutrino detectors are all continental crust based detectors; due to their proximity to the enriched reservoir, up to $75\%$ of the total geo-neutrino flux at these sites is predicted to originate from the continental crust. Thus, refining the composition of the continental crust is an essential prerequisite to using geo-neutrinos to probe radiogenic heating in the deep Earth.
\par
\begin{figure}[h!]
\centering
\includegraphics[width=5in]{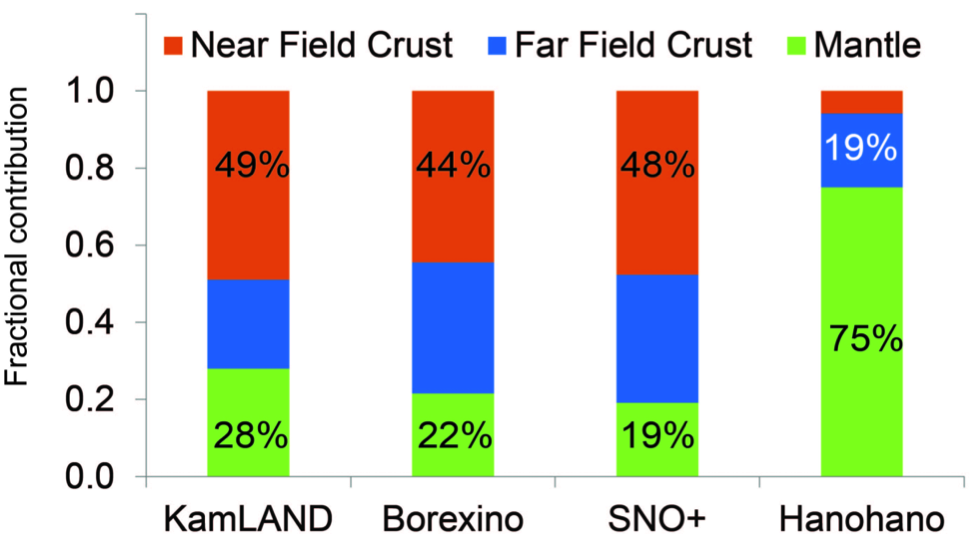}
\caption{\label{Yufig} Predicted fractional geo-neutrino contributions to various detectors from near-field crust (defined as the six closest $2^{\circ}$ by $2^{\circ}$ crust voxels), far-field crust (the rest of crust after removing near-field crust), and assumes one medium Q prediction of the mantle \cite{mcdsun95}.}
\end{figure}

The updated global scale reference model \cite{huang13} combines three existing geophysical models of the global crust to derive the average crustal structure and associated uncertainty. This model includes the previous upper continental crustal composition model \cite{rudgao03} and focuses on refining the chemical composition of the deep continental crust on the basis of compiled compositional databases of amphibolite- and granulite-facies rocks (representative samples from the middle and lower continental crust), along with using seismic velocity data to constrain the lithology in these layers. The U and Th abundances in the databases exhibit log-normal distributions, i.e., the logarithms of their abundances are normally distributed. Monte Carlo simulation was used to combine the uncertainties on physical structure of the continental crust, seismic velocity and U and Th abundances. The dominant source of uncertainty on the predicted geo-neutrino signal from continental crust is the distribution of U and Th abundances. Figure \ref{Yufig} shows the predictions of the geo-neutrino contributions to different detectors from different reservoirs based on the updated model.

\par
The predicted continental crust geo-neutrino signals at the three detectors using the global reference model yield larger uncertainty than is permissible to distinguish various BSE compositional models (section 2). These detection sites have limited sensitivity to the mantle geo-neutrino signal due to the uncertainty on the predicted crust signal. The regional crust (i.e., 500 km radius circle) around the detection site may contribute more than one-half of the total crust signal \citep[e.g.][] {chen06, enomoto}. Thus, independent evaluation of the regional crust signal by investigating the local geology, geophysics, and geochemistry near the detection site is desired. Local geology maps provide higher resolution constraints on surface exposure of various lithologies. The physical structure of the regional crust is typically obtained by combining results from refraction and reflection surveys, drilling wells, and structural geology research. The distributions of U and Th abundances in dominant lithologic units in the regional crust are evaluated from analyses of representative samples. Such regional crust models around KamLAND and Borexino have been made to predict the regional crust signals at these two sites \cite{enomoto, coltorti}. The geo-neutrino signals from the regional crust and the rest of crust are summed to predict the total crust signal; and their uncertainties, assumed to be independent, are added in quadrature.

\section{Deep Mantle Features}
Structure in the lowermost mantle is dominated by a pair of large ($\sim5000$ km across, and extending $>1000$ km from the CMB) low shear velocity zones (LLSVPs) \citep[e.g.][]{garnero2008, dziewonski2010}, one centered beneath the Pacific basin and the other beneath Africa and the Atlantic. Since they were first glimpsed in the very first global tomographic models \citep[eg.][]{adam77}, a number of unusual characteristics of the LLSVPs have been identified. Within the LLSVPs, bulk sound and shear wave speeds may be anti-correlated \cite{su97, masters2000, hernlund2008}. Steep lateral gradients in shear wave velocity appear to bound both LLSVPs \cite{wen01, to05, he06, take08, he09}. Modeling of the splitting of Earth's free oscillations suggests that the LLSVPs may be denser than the surrounding mantle \citep[e.g.][]{ishii99, resovsky03}, though their excess density is likely to be less than 1.5\% \citep{kuo02}. These seismic characteristics suggest that LLSVPs are compositionally distinct from ambient lower mantle, motivating speculation that they represent accumulation of subducted oceanic crust \citep[e.g.][]{christensen1994}, or primordial thermo-chemical piles formed by segregation of dense melts \citep[e.g.][]{lee2010upside}, or as a residue of basal magma crystallization \citep{labrosse2007}. It has also been suggested that the LLSVPs may be enriched in heat producing elements, and may represent the missing heat reservoir that reconciles geophysical and geochemical estimates of Earth's heat budget \citep{kellogg1999}.
\par
The core-mantle boundary region also hosts ultra low velocity zones (ULVZs), which are more than an order of magnitude smaller than LLSVPs, typically extending $\sim100$ km across and $\sim10$ km up from the CMB. Their large shear wave velocity reductions ($>20\%$ slower compared to ambient mantle) have been interpreted as evidence for the presence of partial melt \cite{will96, lay04} or high iron content \cite{mao06, wicks10}. The ULVZs also appear to preferentially occur near the boundaries of the LLSVPs, suggesting a dynamical connection between them, the LLSVPs, and mantle circulation \cite{mcnamara2010}.
\par
Recently, structures with a characteristic spatial scale intermediate between that of the LLSVPs and ULVZs have been detected in the core-mantle boundary region. Cluster analysis of tomographic models \cite{ved12} detects the existence of an isolated structure $\sim1000$ km across and $\sim400$ km tall (the Perm Anomaly). Waveform modeling using the coupled spectral element method \cite{capd03} of shear waves passing through the Perm Anomaly revealed it to have LSVP-like seismic characteristics. Waveform modeling also revealed the presence of two unusually large (`mega') ULVZs, one centered beneath Hawaii \cite{cott12} and the other approximately beneath Tonga \cite{thorne13}. These discoveries have revealed the existence of `mesoscale' structures at the base of the mantle, whose origin and composition remain to be explored.
\par
Taken together, a seismological picture of the Earth's lower mantle emerges, in which structures spanning orders of magnitude in spatial scale bear evidence of substantial variations in temperature and composition. Additionally, the hypotheses formulated to explain the origin and nature of these structures imply different variations in major and trace element compositions, and therefore also potentially different geo-neutrino fluxes. Thus, seismically imaged structures in the lower mantle may produce signals detectable by geo-neutrino observatories \cite{sramek13}, presenting the prospect of using geo-neutrino science to shed light on the processes giving rise to the main seismically-imaged structures in the Earth's deep interior. Conversely, the variations in geo-neutrino flux observed at the surface may be influenced by spatial variations due to a deep mantle signal; correcting for these variations, if present, would be necessary for drawing inferences on average mantle abundances of heat producing elements (K, U, Th).

\section{Mantle Geo-neutrino Flux Resolution Strategies}
The present quest of constraining radiogenic heat production in the mantle requires subtracting the crust flux from the total flux of geo-neutrinos measured at the surface. Models predict the crust flux dominates the total geo-neutrino signal at continental locations (70-80 \% of total signal; Figure \ref{Yufig}: KamLAND, Borexino, SNO+). Refinement of the predicted crust flux, as discussed in section 3, is crucial to reducing the uncertainty of the estimated mantle flux. Combining measurements from various continental detection sites reduces the uncertainty.

\begin{figure}[h!]
\centering
\includegraphics[width=5in]{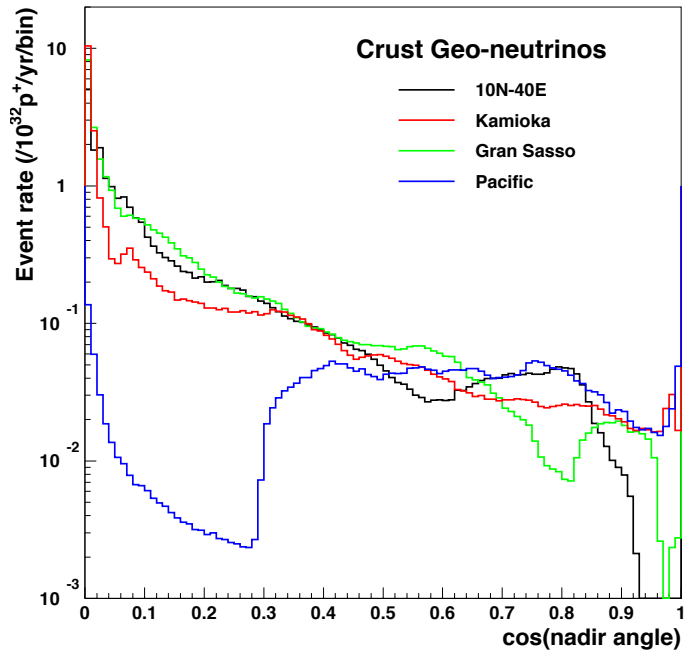}
\caption{\label{crdir} The distribution of the predicted geo-neutrino signal from the crust at existing and selected measurement sites as a function of the cosine of the nadir angle.}
\end{figure}

\par
One possibility to improve sensitivity to mantle geo-neutrinos is to measure the flux in oceanic areas far away from continents. This is because the oceanic crust is thinner ($\sim7$\,km) as opposed to average continental crust thickness of $\sim35$\,km) and about a factor of 10 less enriched in potassium, thorium and uranium than continental crust \cite{huang13}. Therefore the crust flux at a deep oceanic location is smaller compared to a continental location. For most models the total geo-neutrino signal at a deep oceanic location is dominated by flux from the mantle (Figure \ref{Yufig}: Hanohano). An oceanic measurement could sufficiently reduce the uncertainty on mantle geo-neutrino flux to successfully discriminate between Earth models. A multi-site oceanic measurement could be used to test the hypothesis of large scale compositionally distinct structures in the deep mantle, discussed in section 4 \cite{sramek13}. An ocean going detector Hanohano is being planned \cite{dye06, learned:2008}.
\par
An alternative to measuring far away from continents requires the ability to distinguish between crust and mantle signal at the instrument level \citep[e.g.,][]{dye12}. This could be achieved with a directional detector capable of resolving the angular distribution of incoming anti-neutrinos \cite{tanaka}. While crust geo-neutrinos would sharply peak at near-horizontal incident angles (Figure \ref{crdir}), mantle geo-neutrinos would be distributed over a range of angles below the horizontal \citep[e.g.,][]{dye10, mareschal}.

\section{Conclusions}
Geo-neutrino measurements are beginning to resolve the flux from the mantle, providing the first direct estimates of radiogenic heating in this remote reservoir. The level of mantle heating discriminates between potential Earth composition models with implications for the origin and thermal evolution of our planet. Although the uncertainty in the present estimate is quite large, additional exposure from existing and new observatories promises improvement. Results are enhanced by refinements to the predicted crust flux and knowledge of compositional heterogeneities in seismically resolved deep mantle features. Whereas present observatories record only the energy of the geo-neutrinos, new techniques for measuring geo-neutrino direction would strengthen the significance of the measurements.
\section*{Acknowledgments}
This work was supported in part by National Science Foundation grants through the Cooperative Studies of the EarthÕs Deep Interior program (EAR 0855838 and EAR 1068097) and by Hawaii Pacific University grants through the Scholarly Endeavors Program.
\label{}


%% References


\bibliographystyle{elsarticle-num}
\bibliography{Taup_bib}






% \end{thebibliography}

\end{document}